\documentclass[preprint,floatfix,nofootinbib,superscriptaddress]{elsarticle}

\usepackage{graphicx,subfig,epsfig,epsf,color}
\usepackage{amssymb,amsmath}
\usepackage{slashed}
\usepackage{feynmf}
\usepackage{array}
\usepackage{hyperref}

\newcolumntype{C}[1]{>{\centering\let\newline\\\arraybackslash\hspace{0pt}}m{#1}}

\newcommand{\beq}{\begin{equation}}
\newcommand{\eeq}{\end{equation}}
\newcommand{\bea}{\begin{eqnarray}}
\newcommand{\eea}{\end{eqnarray}}
\newcommand{\beas}{\begin{eqnarray*}}
\newcommand{\eeas}{\end{eqnarray*}}
\newcommand{\bcr}{\begin{center}}

\def\Re{{\cal R \mskip-4mu \lower.1ex \hbox{\it e}\,}}
\def\Im{{\cal I \mskip-5mu \lower.1ex \hbox{\it m}\,}}

\def\etal{{\it et al.}}

\def\tev{\,{\ifmmode\mathrm {TeV}\else TeV\fi}}
\def\gev{\,{\ifmmode\mathrm {GeV}\else GeV\fi}}
\def\mev{\,{\ifmmode\mathrm {MeV}\else MeV\fi}}
\def\to{\rightarrow}

\begin{document}

\def\issue(#1,#2,#3){#1 (#3) #2} 
\def\APP(#1,#2,#3){Acta Phys.\ Polon.\ \issue(#1,#2,#3)}
\def\ARNPS(#1,#2,#3){Ann.\ Rev.\ Nucl.\ Part.\ Sci.\ \issue(#1,#2,#3)}
\def\CPC(#1,#2,#3){comp.\ Phys.\ comm.\ \issue(#1,#2,#3)}
\def\CIP(#1,#2,#3){comput.\ Phys.\ \issue(#1,#2,#3)}
\def\EPJC(#1,#2,#3){Eur.\ Phys.\ J.\ C\ \issue(#1,#2,#3)}
\def\EPJD(#1,#2,#3){Eur.\ Phys.\ J. Direct\ C\ \issue(#1,#2,#3)}
\def\IEEETNS(#1,#2,#3){IEEE Trans.\ Nucl.\ Sci.\ \issue(#1,#2,#3)}
\def\IJMP(#1,#2,#3){Int.\ J.\ Mod.\ Phys. \issue(#1,#2,#3)}
\def\JHEP(#1,#2,#3){J.\ High Energy Physics \issue(#1,#2,#3)}
\def\JPG(#1,#2,#3){J.\ Phys.\ G \issue(#1,#2,#3)}
\def\MPL(#1,#2,#3){Mod.\ Phys.\ Lett.\ \issue(#1,#2,#3)}
\def\NP(#1,#2,#3){Nucl.\ Phys.\ \issue(#1,#2,#3)}
\def\NIM(#1,#2,#3){Nucl.\ Instrum.\ Meth.\ \issue(#1,#2,#3)}
\def\PL(#1,#2,#3){Phys.\ Lett.\ \issue(#1,#2,#3)}
\def\PRD(#1,#2,#3){Phys.\ Rev.\ D \issue(#1,#2,#3)}
\def\PRL(#1,#2,#3){Phys.\ Rev.\ Lett.\ \issue(#1,#2,#3)}
\def\PTP(#1,#2,#3){Progs.\ Theo.\ Phys. \ \issue(#1,#2,#3)}
\def\RMP(#1,#2,#3){Rev.\ Mod.\ Phys.\ \issue(#1,#2,#3)}
\def\SJNP(#1,#2,#3){Sov.\ J. Nucl.\ Phys.\ \issue(#1,#2,#3)}

\bibliographystyle{elsarticle}

\title{An Extensive Study of Bose-Einstein Condensation in Liquid Helium using Tsallis Statistics}

\author[bits]{Atanu~Guha\corref{cor1}}
\ead{am.atanu@gmail.com/ p20140401@goa.bits-pilani.ac.in}

\author[bits]{Prasanta~Kumar~Das\corref{cor1}}
\ead{pdas@goa.bits-pilani.ac.in}

\cortext[cor1]{Corresponding author}
\cortext[cor2]{Co-author}

\address[bits]{Birla Institute of Technology and Science-Pilani, Department of Physics, 
Goa campus, NH-17B, Zuarinagar, Goa-403726, India}

\date{\today}

\begin{abstract} 
Realistic scenario can be represented by general canonical ensemble way better than the  ideal one, with proper parameter sets involved. We study the Bose-Einstein condensation phenomena of liquid helium within the framework 
of Tsallis statistics. With a comparatively high value of the deformation parameter $q(\sim 1.4)$, 
the theoretically calculated value of the critical temperature($T_c$) of the phase transition of liquid helium is found to agree with the experimentally determined value ($T_c = 2.17~\rm{K}$), although they differs from each other for $q=1$ (undeformed scenario). This throws a light on the understanding of the phenomenon and connects temperature fluctuation(non-equilibrium conditions) with the interactions between atoms qualitatively. More interactions between atoms give rise to more non-equilibrium conditions which is as expected.

\noindent {{\bf Keywords}: Tsallis statistics, Bose-Einstein condensation, liquid helium.}
\end{abstract}

\maketitle

\section{Introduction}

 Statistical mechanics, an important tool in Theoretical Physics, has been successfully used 
not only in different branch of physics (e.g. condensed matter physics, high energy physics, Astrophysics
 etc.), but also found to be useful in understanding share price dynamics, traffic control dynamics,
 etc), hydroclimatic fluctuations, random networks etc. The results predicted by the Statistical Mechanics have been found to be in 
 good agreement with the experiments. 
 
 Several attempts have been made to generalize this statistical mechanics in recent years 
 \cite{Tsallis1,Sumiyoshi,Ugur,Tsallis2, cohen, Tsallis3} and it (popularly known as superstatistics or $q$-generalized(Tsallis) 
 statistics, where $q$ is the deformation parameter) has already been applied to a wide range of complex systems, e.g., hydrodynamic turbulence, defect turbulence, 
 share price dynamics, random matrix theory, random networks, wind velocity fluctuations, 
 hydroclimatic fluctuations, the statistics of train departure delays and models of the 
 metastatic cascade in cancerous systems \cite{Tsallis4, Tsallis5, Plastino, Plastino2, Plastino3, Fevzi2, Fevzi3, Wilk, Rajagopal1}. In recent times many authors studied the thermostatic properties of different kind of physical systems(which are more complex than an ideal gas system) like self-gravitating stellar system, Levy flight random diffusion, the galaxy model of the generalized Freeman disk, the electron-plasma 2-$D$ turbulence, the cosmic background radiation, correlated themes, the linear response theory, solar neutrinos, thermalization of electron$-$photon systems etc \cite{Salazar, Ion, Buiatti, Prato, Wang, Levy, Boghosian, Plastino20, Hamity}.

 This approach deals with the fluctuation parameter $q$ which corresponds to the degree of 
 the temperature fluctuation effect to the concerned system. Here we can treat 
 our normal Boltzmann-Gibbs statistics as a special case of this generalized one, where 
 temperature fluctuation effects are negligible, corresponds to $q=1.0$. 
 More deviation of $q$ from the value $1.0$ denotes a system with more fluctuating 
 temperature. Various works related to this $q$-generalized or Tsallis 
 statistics have been reported in different phenomena
 \cite{Ugur, cbeck1,cbeck2,cbeck3,cbeck4, Tsallis6, Tsallis7, Lyra, Alemany, Lenzi10, Plastino30, Plastino40, Torres, Rajagopal2, Kaniadakis, Koponen, sisto, Atanu}.
 

\section{Connection between entropy and microstates in Tsallis statistics}
 A simple connection between the entropy($s$) and the microstates($\Omega$) of a system
 can be easily derived as $ s=k_B \ln \Omega $, where one assumes that the entropy($s$) is 
additive,while the number of microsates($\Omega$) is multiplicative. 
 
 A more general connection between $s$ and $\Omega$ can be shown \cite{Atanu1, Atanu2} to be equal to 
\bea
s_q=k_B \ln_q \Omega
\eea
where the generalized log function($\ln_q \Omega$) is defined as
\bea
\ln_q \Omega = \frac{\Omega^{1-q}-1}{1-q}.
\label{eqn:lnq}
\eea
Consequently the generalized exponential function becomes
\bea
e_q^x=\left[1+(1-q) x \right]^\frac{1}{1-q}.
\label{eqn:expq}
\eea

Therefore $q$-modified Shanon entropy takes the following form
\bea
s_q=-k_B <\ln_q p_i> = k \frac{\sum p_i^q -1}{1-q}
\eea

Extremizing $s_q$ subject to suitable constraints yields more general canonical ensembles(see \ref{entropy_optimization}),
where the probability to observe a microstate with energy $\epsilon_i$ is given by: \cite{Tsallis1, Tsallis12, oikonomou}  
\bea
p_i = \frac{e_q^{- \beta' \epsilon_i}}{Z_q} = \frac{1}{Z_q} \left[ 1 - (1-q) \beta' \epsilon_i \right]^{\frac{1}{1-q}}
\eea
with partition function $z_q$ and inverse temperature parameter $\beta=\frac{1}{k_B T}$. Also $\beta'$ is the $q$-modified quantity and is given by \cite{Tsallis1, Tsallis12} 
\bea
\beta'=\frac{\beta}{\sum_i p_i^q+(1-q) \beta u_q}=\frac{\beta}{Z_q^{1-q}+(1-q) \beta u_q}
\eea 

 with $q$-generalized average energy 
\bea
u_q=\frac{\sum_i \epsilon_i p_i^q}{\sum_i p_i^q}
\eea 
 
 In the limit of small deformation approximation(i.e., small $\mid 1-q \mid $), $ \beta' \to \beta $


\section{Bose-Einstein Condensation of liquid $He$ in the framework of $q$-generalized Tsallis Statistics}

\label{be}
 Whenever a system is subjected to the temperature fluctuation, the non$-$equilibrium 
 generalized statistical mechanics plays a crucial role. 
 If the temperature fluctuation effect is not negligible enough to disclose itself, 
 then it is expected to observe some deviation from the ideal phenomena. Here we study 
 one such phenomena i.e. Bose-Einstein condensation phenomena in Liquid Helium.

 The Pauli-Exclusion principle forbids any two fermions to sit at the lowest (or any other 
value) energy states, while no such principle forbids particles with integral spins to 
occupy the same quantum states. This gives rise many interesting properties at low 
temperature and the Bose-Einstein condensation is one of them. With zero spin, a $He^4_2$ 
atom is a boson and does not obey the Pauli-Exclusion principle. In 1911, Kamerlingh Onnes first discovered liquid Helium($He^4$) at a temperature of $4.2~\rm{K}$ \cite{Kamerlingh}. 
While plotting the specific heat 
as a function of the temperature for liquid helium $He^4_2$, Keesom and Clausius 
in 1932 \cite{Keesom}, first found a discontinuity in the specific heat at a temperature 
$T = 2.17~\rm{K}$ (called the ``critical temperature'') and the specific heat jumped to 
a large value - a phase transition in which liquid helium goes from its normal phase 
(i.e. liquid helium phase I) to superfluid phase (liquid helium phase II). 

 For the liquid helium the theoretically predicted value was $ \sim 3.1 K$, whereas 
 experiments suggest that the superfluid state of liquid helium has been obtained
 near $ \sim 2.17 K$. This happens because the interactions between the atoms are too strong. 
 Only $8 \%$ of atoms are in the ground state near absolute zero, rather than the $ 100 \%$ 
 of a true condensate \cite{london, royal, bradley, dale}. 
 
To study BE-condensation, let us first write down the grand canonical partition function, which 
in Tsallis statistics, takes the following form:
\bea
\mathcal{Z}_q (T,V,\mu) = \sum_{\left\lbrace n_k \right\rbrace=0}^{\infty} \exp_q {\left\lbrace -\beta \sum_{k=1}^{\infty} n_k \left( \epsilon_k - \mu \right) \right\rbrace}
\label{zq1}
\eea 
where, $\exp_q (x)$ is the $q$-generalized exponential function, given by Eq.(\ref{eqn:expq}).

\begin{figure}[htbp]
\includegraphics[width=11cm]{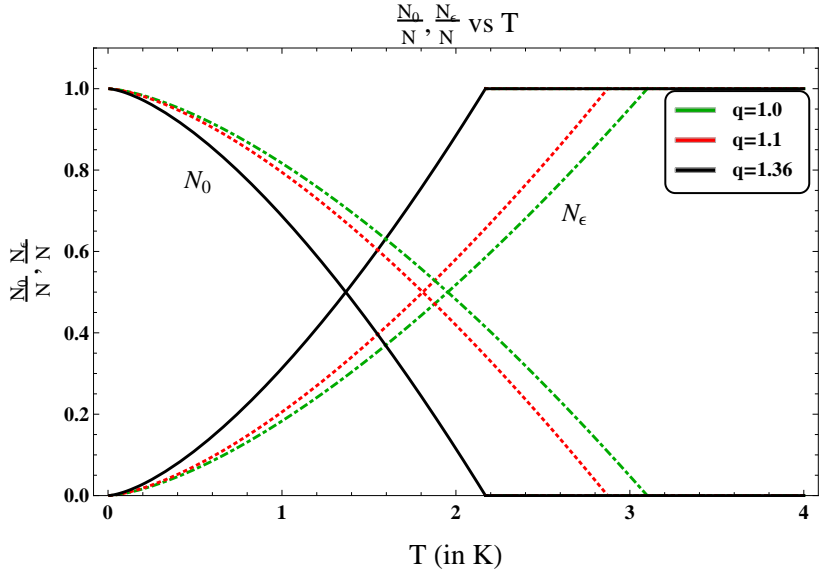}
\caption{\it Plots showing the variation of  $N_0/N$ and $N_{\epsilon}/N$ as a function of $T$  for $q =1.0,~1.1$ and $1.36$, respectively. Plots are for liquid helium with 
$n = 2.2 \times 10^{28}~\rm{Atoms/m^3}$ and mass of helium atom is taken to be $m=6.8 \times 10^{-27}~\rm{kg}$.}
\label{NNexT}
\end{figure}

For small deformation (i.e. negligible temperature fluctuation), we
find (using Eqs. (\ref{smlmul}) and (\ref{smlpwr}); please refer to \ref{weak_deformation})
\bea
\mathcal{Z}_q &=& \sum_{n_1, n_2, \cdots = 0}^{\infty} \left[ \exp_q {\left\lbrace -\beta \left( \epsilon_1 - \mu \right) \right\rbrace} \right]^{n_1} \left[ \exp_q {\left\lbrace -\beta \left( \epsilon_2 - \mu \right) \right\rbrace} \right]^{n_2} \cdots \nonumber \\
&=& \prod_{k=1}^{\infty} \sum_{n_k=0}^{\infty} \left[ \exp_q {\left\lbrace -\beta \left( \epsilon_k - \mu \right) \right\rbrace} \right]^{n_k} \nonumber \\
&=& \prod_{k=1}^{\infty} \frac{1}{1- z_q \exp_q {(- \beta \epsilon_k)}}
\label{zq2}
\eea 
In above $ z_q=\exp_q (\beta \mu)$ ($ \mu $ is the chemical potential) is the $q$-generalized fugacity.
The average number of particle(normalized) in $k$-th state(with energy $\epsilon_k$) 
\bea
<n_k>_q=\frac{\sum_{n_k=0}^{\infty} n_k p_k^q}{\sum_{n_k=0}^{\infty} p_k^q}
\label{nkavg1}
\eea
where the probability distribution $p_k$ is given by
\bea
p_k &=& \frac{1}{\mathcal{Z}_q} \left[ \exp_q {\left\lbrace -\beta(\epsilon_k - \mu) \right\rbrace}\right]^{n_k} \nonumber \\
&=& \frac{1}{\mathcal{Z}_q} z_q^{n_k} \exp_q (-\beta n_k \epsilon_k)
\label{pk}
\eea
Substituting Eq.(\ref{pk}) into Eq.(\ref{nkavg1}) and simplifying further we get
\bea
<n_k>_q=\frac{1}{\left( z_q e_q^{-\beta \epsilon_k}\right)^{-q}-1}
\label{nkavgfnl}
\eea
For $q=1$ Eq.(\ref{nkavgfnl}) exactly replicates the undeformed scenario, which states
\bea
<n_k>=\frac{1}{ z^{-1} e^{\beta \epsilon_k}-1}
\eea
Now the total number of particles (including the ground state)
\bea
N &=& \sum_k <n_k>_q \nonumber \\
&=& \frac{1}{z_q^{-q}-1} + \sum_{k \neq 0} \frac{1}{\left( z_q e_q^{-\beta \epsilon_k}\right)^{-q}-1} \nonumber \\
&=& N_0 + N_{\epsilon}
\eea
with 
$N_0=\frac{1}{z_q^{-q}-1}$
and
$N_{\epsilon} = \frac{V}{\lambda^3} ~ ga_{3/2}(z_q)$, being the number of particles in the ground state and in the excited states.
Here $\lambda(=h/ \sqrt{2 \pi m k_B T}~)$ is the thermal de-Broglie wavelength  
and $ga_{3/2}(z_q)$ is the $q$-generalized polylog function(Bose integral) of the first kind, given by
\bea
ga_{3/2}(z_q)=\frac{1}{\Gamma(3/2)} \int_0^{\infty} \frac{dx ~ x^{3/2-1} }{\left( z_q e_q^{-x}\right)^{-q}-1}
\eea
$ x= \beta \epsilon $ is the dimensionless quantity. Using the expressions in $q$-generalized Tsallis scenario we get the characteristic(i.e. critical) temperature for Bose-Einstein condensation \cite{Patharia}
as follows
\bea
T_c=\frac{h^2}{2 \pi m k_B}\left[\frac{n}{ga_{3/2}(z_q=1)}\right]^{2/3}
\label{def:tc}
\eea
where, $m$ denotes the mass the of the particle species concerned and $n$, the number of 
the particles per unit volume(i.e. number density(=$N/V$)) respectively. It clearly shows the 
dependence of $T_c$ on the deformation parameter $q$. Below in Fig.[\ref{tcq}], we have 
shown the dependence of the Bose-Einstein condensation temperature ($T_c$)  
on the deformation parameter $q$ for liquid helium\cite{Patharia} (with $n = 2.2 \times 10^{28}~\rm{Atoms/m^3}$ and 
$m = m_{He}=6.8 \times 10^{-27}~\rm{kg}$). 
\begin{figure}[htbp]
\includegraphics[width=11cm]{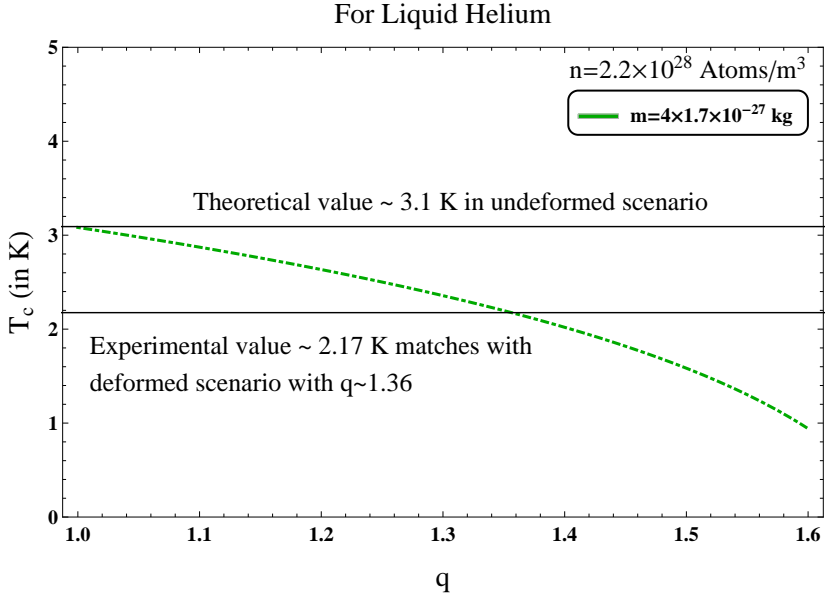}
\caption{\it The dependency of the condensation temperature($T_c$) for liquid helium on the 
deformation parameter($q$) is shown.}
\label{tcq}
\end{figure}
The upper horizontal curve corresponds to $T_c = 3.1 ~K$ in undeformed 
scenario($q=1$), while the lower horizontal curve corresponds to 
$T_c = 2.17~K$ (the experimental data for liquid hydrogen). The difference 
between the theoretical prediction and the experimental value of $T_c$ for 
liquid helium can be explained using Tsallis statistics. 
From the figure we see that  the helium condensation temperature $T_c$ as predicted by the Tsallis statistics 
agrees with the experimental value corresponding to $q \sim 1.36$, thereby signifies the 
importance of  deformed statistics which can explain the difference between the theory (undeformed value) and the experiment.
In Fig.[\ref{NNexT}], we have plotted $N_0/N$ and $N_{\epsilon}/N$ as a function of $T$ 
corresponding to $q = 1.0,~1.1$ and $1.36$ for liquid helium\cite{Patharia}.
From the figure(Fig. \ref{tcq}), we see that as $q$ increases, the critical temperature($T_c$) of the Bose-Einstein condensation of liquid helium decreases and eventually  
matches with the experimental value at $T_c = T_c^{exp}= 2.17~\rm{K}$ for the deformation parameter $q=1.36$.
\subsection{Specific heat variation and BE condensation}
From Eq.{\ref{zq2}}, we find the partition function after simplification, 
\bea
\ln \mathcal{Z}_q &=& \ln \left[ \prod_{k=1}^{\infty} \frac{1}{1- z_q \exp_q {(- \beta \epsilon_k)}}  \right] \nonumber \\
&=& \ln (1-z_q)+\frac{V}{\lambda^3} ~ gb_{5/2}(z_q),
\eea
where $gb_{5/2}(z_q)$ is the $q$-generalized polylog function(Bose integral) of the second kind, given by
\bea
gb_{5/2}(z_q)=\frac{1}{\Gamma(5/2)} \int_0^{\infty} \frac{dx ~ x^{5/2-1} }{z_q^{-1} \left( e_q^{-x}\right)^{-q}-\left( e_q^{-x}\right)^{1-q}}
\eea
The internal energy $U_q$($q$-generalized internal energy) is given by \cite{Tsallis1, Sumiyoshi}
\bea
U_q = -\frac{\partial}{\partial \beta} \ln_q \mathcal{Z}_q  = -\frac{\partial}{\partial \beta} \ln_q e^{\ln \mathcal{Z}_q}
\eea
 where the $q$-generalized logarithm is defined by Eq.(\ref{eqn:lnq}). The normalized 
 $q$-generalized internal energy is defined as \cite{Tsallis1, Swamy, guo}
\bea
<U_q>=\frac{U_q}{\sum_i p_i^q} = \frac{3}{2} \frac{V}{\lambda^3} k_B T ~ gb_{5/2}(z_q)
\label{eqn:int_enrgy}
\eea
In BE condensation phase(i.e., $ T<T_c $), the fugacity $ z_q=1 $. So in this phase the molar 
specific heat capacity of the system at constant volume is given by
\bea
C_V=\left[\frac{\partial <U_q>}{\partial T}\right]_{N,V}=\frac{3}{2} V k_B ~gb_{5/2}(z_q=1) \frac{\partial}{\partial T} \left(\frac{T}{\lambda^3}\right)
\eea
Now using the fact that 
$ \frac{\partial}{\partial T} \left(\frac{T}{\lambda^3}\right)= \frac{5}{2} \frac{1}{\lambda^3} $, 
we find 
\bea
\frac{C_V}{N k_B}=\frac{15}{4} \frac{V}{N} \frac{1}{\lambda^3} ~ gb_{5/2}(z_q=1) \propto T^{3/2}
\label{cvnk2}
\eea
with $ T<T_c $ (in condensation phase). For $ T>T_c $, $z_q<1$ and $N_0 \approx 0$.
\bea
\therefore N &\approx & \frac{V}{\lambda^3}~ ga_{3/2}(z_q) \implies \frac{V}{\lambda^3} = \frac{N}{ga_{3/2}(z_q)}
\label{reln}
\eea

From Eq.(\ref{eqn:int_enrgy})
\bea
<U_q> &=& \frac{3}{2} \frac{N}{ga_{3/2}(z_q)} k_B T ~ gb_{5/2}(z_q) = \frac{3}{2} N k_B T ~\frac{gb_{5/2}(z_q)}{ga_{3/2}(z_q)}
\eea  
and
\bea
\frac{C_V}{N k_B} &=& \frac{1}{N k_B} \left[\frac{\partial <U_q>}{\partial T}\right]_{N,V} \nonumber \\
&=& \frac{3}{2} \left[ \frac{gb_{5/2}(z_q)}{ga_{3/2}(z_q)} + T \frac{\partial}{\partial T} \left\lbrace \frac{gb_{5/2}(z_q)}{ga_{3/2}(z_q)} \right\rbrace\right] \nonumber \\
&=& \frac{3}{2} \left[ \frac{gb_{5/2}(z_q)}{ga_{3/2}(z_q)} + T \frac{\partial z_q}{\partial T} \frac{\partial}{\partial z_q} \left\lbrace \frac{gb_{5/2}(z_q)}{ga_{3/2}(z_q)} \right\rbrace\right]
\label{cvnk1}
\eea
 
Now using $\frac{\partial}{\partial T} \left( \lambda^3 \right)= - \frac{3}{2} \frac{\lambda^3}{T} $ and Eq.(\ref{reln}) we get 
\bea
\frac{\partial z_q}{\partial T}= - \frac{3}{2} \frac{1}{T} \frac{ga_{3/2}(z_q)}{ga'_{3/2}(z_q)}
\eea
where, $ga'_{3/2}(z_q)$ denotes the derivative of $ga_{3/2}(z_q)$ with respect to $z_q$.
Putting this back on Eq.(\ref{cvnk1}) the expression for the molar specific heat capacity per unit volume becomes
\bea
\frac{C_V}{N k_B} &=& \frac{3}{2} \left[ \frac{gb_{5/2}(z_q)}{ga_{3/2}(z_q)} - \frac{3}{2} \frac{ga_{3/2}(z_q)}{ga'_{3/2}(z_q)} \frac{\partial}{\partial z_q} \left\lbrace \frac{gb_{5/2}(z_q)}{ga_{3/2}(z_q)} \right\rbrace\right]
\eea
Simplifying further we get
\bea
\frac{C_V}{N k_B} &=& \frac{15}{4}  \frac{gb_{5/2}(z_q)}{ga_{3/2}(z_q)} - \frac{9}{4}  \frac{gb'_{5/2}(z_q)}{ga'_{3/2}(z_q)}
\label{cvnkfnl}
\eea
Eq.(\ref{cvnkfnl}) is valid for $T>T_c$. For the BE condensation phase in which $T<T_c$, 
the  simplified form of Eq.(\ref{cvnk2}) becomes
\bea
\frac{C_V}{N k_B}=\frac{15}{4} \frac{gb_{5/2}(z_q=1)}{ga_{3/2}(z_q=1)} \left(\frac{T}{T_c}\right)^{3/2}
\eea
Finally, using the definition of $T_c$, Eq.(\ref{def:tc}) we get
\bea
\frac{1}{\lambda^3}= \left(\frac{T}{T_c}\right)^{3/2} \frac{N/V}{ga_{3/2}(z_q=1)}
\eea
\begin{figure}[htbp]
\includegraphics[width=10cm]{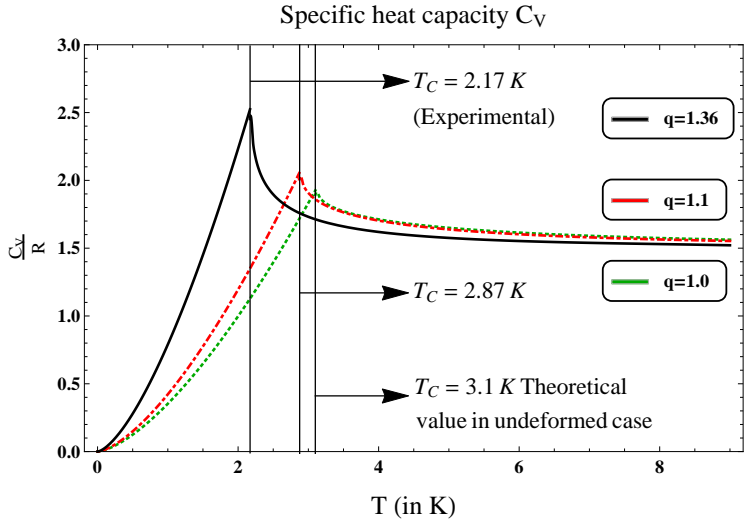}
\caption{\it The specific heat $C_V/R$ is plotted against $T$ for liquid helium. 
The plots show the phase transition of liquid helium from its normal phase to it superfluid phase. The plots are shown for $q=1.0,~q=1.1$ and $1.36$, respectively. }
\label{cvbeq}
\end{figure}
In Fig.[\ref{cvbeq}], we have plotted the specific heat $C_V/R$ (of liquid helium) as a 
function of $T$ corresponding to the different values of  the deformation parameter $q$. 
We vary $q$ varies from $q=1$ to $q=1.36$ for liquid helium\cite{Patharia} with $n = 2.2 \times 10^{28}~\rm{Atoms/m^3}$. 
The discontinuity (at the peak) in $C_V/R$ vs $T$ curve corresponds to the phase 
transition i.e. transition from normal phase(phase I) to superfluid phase(phase II). 
We see that as $q$ increases from $1.0$ to $1.36$, the transition temperature $T_c$ changes from 
$3.1~\rm{K}$ to $2.17~\rm{K}$ which is the experimentally determined value. 

 A significant amount of work has already been done to observe the effect of generalized Tsallis statistics on the phenomena of Bose-Einstein condensation \cite{Vorberg, Sukanya, Ou, Chen}. Ou \etal studied the thermostatic properties of a $q$-generalized Bose system trapped in an $n$-dimensional harmonic oscillator potential \cite{Ou} whereas Chen \etal investigated $q$-generalized Bose-Einstein condensation based on Tsallis entropy \cite{Chen}. Both of the above mentioned works deal with a $D$-dimensional $q$-generalized ideal boson system with the general energy spectrum
 \bea
 \varepsilon=ap^s
 \eea 
 with two positive constants $a$ and $s$. For non-relativistic particle system $a=\frac{1}{2m}$ and $s=2$(where $m$ is the mass of the concerned particle) whereas for relativistic system, $s=1$ and $a=c$(the speed of light). 
 
 For $q>1$ they Ou \etal and Chen \etal obtained $q$-generalized Bose integral as follows \cite{Ou, Chen}
 
\bea
g_{q,n}(z_q)=\sum_{j=1}^{\infty} \frac{z_q^{j-(1-q)n}}{(q-1)^n} \frac{\Gamma(j/(q-1)-n)}{\Gamma(j/(q-1))}
\eea 

Consequently the $q$-generaized Riemann Zeta function was defined as
\bea
\zeta_q(n)=\sum_{j=1}^{\infty} \frac{1}{(q-1)^n} \frac{\Gamma(j/(q-1)-n)}{\Gamma(j/(q-1))}, ~~ q>1
\eea

with the interesting fact that
\bea
\lim_{q \to 1} \zeta_q(n) = \zeta(n)
\eea
 Using these facts we can estimate the characteristic(i.e. critical) temperature for Bose-Einstein condensation in $q$-generalized Tsallis scenario which can be expressed as a simple relation now \cite{Ou, Chen}
\bea
\frac{T_{q,c}}{T_c}= \left\lbrace \frac{\zeta(D/s)}{\zeta_q(D/s)} \right\rbrace^{s/D}
\label{ou_tc}
\eea 
 where $T_{q,c}$ and $T_c$ are the Bose-Einstein condensation temperature for $q>1$ and $q=1$ respectively. For a 3 dimensional non-relativistic particle system(as considered in our present work Sec. \ref{be}) $s=2$ and $D=3$. In that case, Eq.(\ref{ou_tc}) becomes
\bea
T_{q,c}= T_c \left\lbrace \frac{\zeta(3/2)}{\zeta_q(3/2)} \right\rbrace^{2/3}
\label{chen_tc}
\eea 
 Now we know that, for $q=1$, $T_c=3.1$. Putting $q=1.36$ into the Eq.(\ref{chen_tc}) yields $T_{q,c}=2.178$ which is in nice agreement with the value of the condensation temperature obtained using Eq.(\ref{def:tc}) for $q=1.36$. Also the most important fact is that for $q=1.36$, both of the Eqs.(\ref{chen_tc}) and (\ref{def:tc}) giving us a unique theoretical value(in $q$-generalized Tsallis scenario) of the characteristic(i.e. critical) temperature for Bose-Einstein condensation which also agrees with the experimentally determined one.

\begin{figure}[htbp]
\centering
\includegraphics[width=10cm]{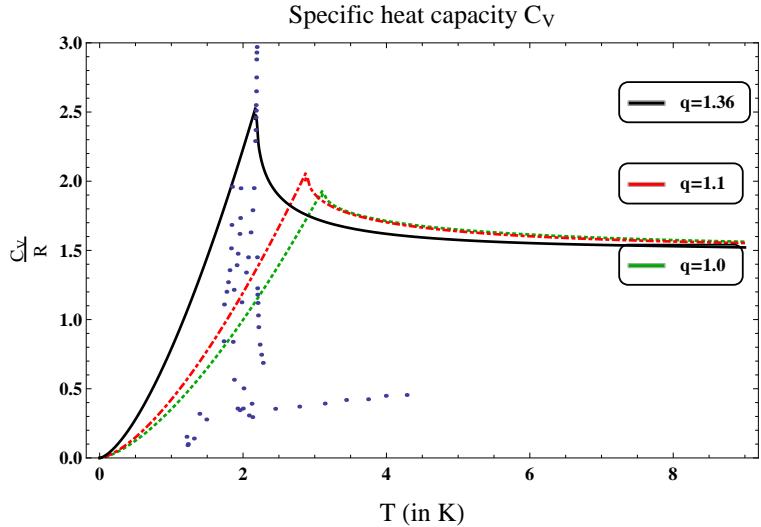}
\caption{\it The specific heat $C_V/R$ is plotted against $T$ for liquid helium. The plots are shown for $q=1.0,~q=1.1$ and $1.36$, respectively with the experimental data points(blue dots) of $C/R$ by Keesom \etal \cite{Keesom2}.}
\label{cvbeqprime}
\end{figure}

 In Fig.[\ref{cvbeqprime}], we have plotted the specific heat $C_V/R$ (of liquid helium) as a function of $T$ corresponding to the different values of  the deformation parameter $q$. Also the experimental data points have been shown which was obtained by W. H. Keesom and A. P. Keesom \cite{Keesom2}.
\section{Conclusion}
We study the Bose-Einstein condensation phenomena within the framework of Tsallis statistics. We find that the critical temperature($T_c$) of the Bose-Einstein condensation depends strongly on the deformation parameter $q$. For $q=1$ (undeformed scenario), we find that the theoretically calculated value of the 
critical temperature ($\sim 3.1~\rm{K}$) differs from the experimentally measured value $\sim 2.17~\rm{K}$. With a relatively high value of the deformation parameter $q$($\sim 1.36$), the theoretical prediction of the critical temperature for liquid helium(which is a highly interacting non-ideal boson system) matches with 
the  experimentally determined one. Here we can consider the deformation parameter $q$ is taking care of the concerned non-equilibrium conditions which arises due to the nearest neighbour interactions among the atoms involved. Consequently the undeformed scenario($q=1.0$, in which we consider an ideal non-interacting bosonic system) failed to explain the discrepancy between the theoretical and the experimentally determined value.

 \section*{ACKNOWLEDGMENTS}
\noindent The authors would like to thank Selvaganapathy J. for useful discussions and valuable suggestions. The work of PKD is supported by the SERB Grant No. EMR/2016/002651. One of the author, Atanu, wants to thank Debashree Sen and Tuhin Malik for advice regarding tools.
\appendix

\section{Indicial properties of $q$-generalized exponential function for small deformation}

 From Eq.(\ref{eqn:expq}), keeping only first order in $(1-q)$,
\bea
e_q^a \cdot e_q^b &=& \left[1+(1-q) a \right]^\frac{1}{1-q} \cdot \left[1+(1-q) b \right]^\frac{1}{1-q} \nonumber \\
&=& \left[1+ (1-q) (a+b) + (1-q)^2 ab \right]^\frac{1}{1-q} \nonumber \\
& \simeq & e_q^{a+b} 
\label{smlmul}
\eea

 Similarly, neglecting higher order terms we get,
\bea
\left(e_q^a \right)^b &=& \left[1+(1-q) a \right]^\frac{b}{1-q} \nonumber \\
&=& \left[1+ (1-q) ab + \frac{b(b-1)}{2!}(1-q)^2 a^2+ \cdots \right]^\frac{1}{1-q} \nonumber \\
& \approx & e_q^{ab} 
\label{smlpwr}
\eea
 
 Though in our present discussion, the $q$-value used to fit the data is $q \approx 1.36 $, the approximation still holds because of the values of $a$ and $b$ to be substituted in Eqs. (\ref{smlmul}) and (\ref{smlpwr}). Eqs. (\ref{zq1}) and (\ref{nkavg1}) are leading to Eqs. (\ref{zq2}) and (\ref{nkavgfnl}), respectively. There we are using the energy of the particle in $k$-th state as $\epsilon_k$ and the $q$-generalized fugacity as $ z_q=\exp_q (\beta \mu)$ with $ \mu $ as the chemical potential. The quantities $\beta \mu$ and $\beta \epsilon_k$ are to be treated as $a$ and $b$ in Eqs. (\ref{smlmul}) and (\ref{smlpwr}) to apply the weak deformation approximation(i.e., $\mid 1-q \mid$). The validity of the approximation holds as the quantities associated with $(1-q)^2$(or higher order of $(1-q)$) is small like $ab, a^2 b, a^2 b^2$ etc. (i.e., $\beta^2 \epsilon_i \epsilon_j, \beta^3 \epsilon_i^2 \epsilon_j, \beta^4 \epsilon_i^2 \epsilon_j^2, \beta^2 \epsilon_i \mu$ etc.). For $q \approx 1.36 $, $(1-q)^2 \approx 0.13$ which is almost $\frac{1}{3}$rd of its first order $(1-q)=0.36$. Clearly we can restrict our consideration up to first order of $\mu$ and $\epsilon_k$ with weak deformation approximation. So even for $q \approx 1.36 $ the weak deformation approximation is reasonably valid as the contribution of the higher order terms in Eqs. (\ref{smlmul}) and (\ref{smlpwr}) becomes negligible. 
 
 \protect\label{weak_deformation}

\section{Constraints and Entropy Optimization in Tsallis Statistics}

 To impose the mean value of a variable in addition to satisfy the following fact
 \bea
 \int_0^{\infty} dx~p(x)=1
 \eea
 
$q$-generalized mean value of a variable $x$ is to be defined as \cite{Tsallis1, Tsallis12}
\bea
<x>_q=\int_0^{\infty} dx~xP(x)=X_q
\eea

whereas, $P(x)$ is the Escort distribution and is defined as
\bea
P(x)=\frac{[p(x)]^q}{\int_0^{\infty} dx'~[p(x')]^q}
\eea

We immediately verify that $P(x)$ is normalized as well
\bea
\int_0^{\infty} dx~P(x)=\frac{\int_0^{\infty} dx~[p(x)]^q}{\int_0^{\infty} dx'~[p(x')]^q}=1
\eea

We can use these facts to optimize the generalized entropy $s_q$. In order to use the Lagrange's undetermined multiplier method to find the optimized distribution we define the following quantity
\bea
\Phi[p]=\frac{1-\int_0^{\infty} dx~[p(x)]^q}{q-1}- \alpha_q \int_0^{\infty} dx~p(x) -\beta_q \frac{\int_0^{\infty} dx~x[p(x)]^q}{\int_0^{\infty} dx~[p(x)]^q}
\eea
with $\alpha_q$ and $\beta_q$ as the Lagrange parameters. Therefore imposing the optimization conditions
\bea
\frac{\partial \Phi(p)}{\partial p}=0
\eea
Simplifying further we get 
\bea
p(x)=\frac{e_q^{-\beta_q \left(x-X_q \right)}}{\int_0^{\infty} dx' ~e_q^{-\beta_q \left(x'-X_q \right)}}
\eea

Now from the following two constraints
\begin{itemize}
\item $ \sum_i p_i=1$ (Norm constraint)
\item $ <\epsilon>_q = \sum_i \epsilon_i P_i= u_q$ (Energy constraint)
 
 with $P_i=\frac{p_i^q}{\sum_j p_j^q}$
\end{itemize}
 we obtain the distribution as follows
\bea
p_i=\frac{e_q^{- \beta_q \left( \epsilon_i - u_q \right)}}{\bar{z}_q}
\label{distribution_q}
\eea 
 
 with $ \bar{z}_q=\sum_i e_q^{- \beta_q \left( \epsilon_i - u_q \right)}$ and $\beta_q=\frac{\beta}{\sum_j p_j^q}$.

 Now 
 \bea
 s_q &=& -k_B <\ln_q p_j> \nonumber \\ 
 &=& k_B \frac{\sum_j p_j^q-1}{1-q} \nonumber \\
  \implies  \sum_j p_j^q &=& 1+ (1-q) \frac{s_q}{k_B}
 \eea

 Also
\bea
s_q = -k_B \ln_q \bar{z}_q= \frac{k_B}{1-q} \left( \bar{z}_q^{1-q} -1 \right)
\eea
 
\bea
\therefore \sum_j p_j^q = 1+ (1-q) \frac{\frac{k_B}{1-q} \left( \bar{z}_q^{1-q} -1 \right)}{k_B} = \bar{z}_q^{1-q} 
\eea
 
So now
\bea
\beta_q=\frac{\beta}{\sum_j p_j^q}= \beta \bar{z}_q^{q-1}
\eea 

 More useful and the convenient form of Eq.(\ref{distribution_q}) for application purpose, is given by \cite{Tsallis1, Tsallis12}
 
\bea
p_i= \frac{e_q^{- \beta' \epsilon_i}}{Z_q}
\eea 

 with $ Z_q=\sum_i e_q^{- \beta' \epsilon_i}$ and $\beta'=\frac{\beta_q}{1+(1-q) \beta_q u_q}=\frac{\beta}{Z_q^{1-q}+(1-q) \beta u_q}$.
 
\protect\label{entropy_optimization}
\section{Converting summations to integrals}
 Using Eq. (\ref{nkavgfnl}) we can calculate the total number of particles (including the ground state) in the following way
\bea
N &=& \sum_k <n_k>_q \nonumber \\
&=& \frac{1}{z_q^{-q}-1} + \sum_{k \neq 0} \frac{1}{\left( z_q e_q^{-\beta \epsilon_k}\right)^{-q}-1} \nonumber \\
&=& N_0 + N_{\epsilon}
\eea
where the number of particles in the ground state
\bea
N_0=\frac{1}{z_q^{-q}-1}
\eea
and the number of particles in the excited states is given by
\bea
N_{\epsilon} = \sum_{k \neq 0} \frac{1}{\left( z_q e_q^{-\beta \epsilon_k}\right)^{-q}-1}
\eea

Converting the above mentioned summation into integral with proper phase-space factor we get

\bea
N_{\epsilon} = \frac{2 \pi V}{h^3} (2 m k_B T)^{\frac{3}{2}} \Gamma(3/2) \frac{1}{\Gamma(3/2)} \int_0^{\infty} \frac{dx ~ x^{1/2} }{\left( z_q e_q^{-x}\right)^{-q}-1} 
\eea
with $ x= \beta \epsilon $ (a dimensionless quantity). The extra factor $(k_B T)^{\frac{3}{2}}$ arises due to this change of variable inside the integral.
Now $\frac{h}{ \sqrt{2 \pi m k_B T}}= \lambda$, the thermal de-Broglie wavelength and $\Gamma(3/2)= \frac{\sqrt{3}}{2}$.
\bea
\therefore  \frac{2 \pi}{h^3} (2 m k_B T)^{\frac{3}{2}} \Gamma(3/2) = \frac{1}{\lambda^3}
\eea
and
\bea
N_{\epsilon} = \frac{V}{\lambda^3} ~ ga_{3/2}(z_q)
\eea
$ga_{3/2}(z_q)$ is the $q$-generalized polylog function(Bose integral) of the first kind, given by
\bea
ga_{3/2}(z_q)=\frac{1}{\Gamma(3/2)} \int_0^{\infty} \frac{dx ~ x^{3/2-1} }{\left( z_q e_q^{-\beta \epsilon_k}\right)^{-q}-1}
\eea

Similarly from Eq.{\ref{zq2}}, we get the following 

\bea
\ln \mathcal{Z}_q &=& \ln \left[ \prod_{k=1}^{\infty} \frac{1}{1- z_q \exp_q {(- \beta \epsilon_k)}}  \right] \nonumber \\
&=& \ln 1 - \ln \left\lbrace 1- z_q e_q^{- \beta \epsilon_1} \right\rbrace -\ln \left\lbrace 1- z_q e_q^{- \beta \epsilon_2} \right\rbrace - \cdots \infty \nonumber \\
&=& \ln (1-z_q)- \sum_{k=0}^{\infty} \ln \left\lbrace 1- z_q e_q^{- \beta \epsilon_k} \right\rbrace
\eea
And after converting the summation into the integral(following the same procedure mentioned above) we get
\bea
\ln \mathcal{Z}_q = \ln (1-z_q)+\frac{V}{\lambda^3} ~ gb_{5/2}(z_q),
\eea
where $gb_{5/2}(z_q)$ is the $q$-generalized polylog function(Bose integral) of the second kind, given by
\bea
gb_{5/2}(z_q)=\frac{1}{\Gamma(5/2)} \int_0^{\infty} \frac{dx ~ x^{5/2-1} }{z_q^{-1} \left( e_q^{-x}\right)^{-q}-\left( e_q^{-x}\right)^{1-q}}
\eea

Intermediate steps:
\bea
- \sum_{k=0}^{\infty} \ln \left\lbrace 1- z_q e_q^{- \beta \epsilon_k} \right\rbrace &=& -\frac{2 \pi V}{h^3} (2m)^{3/2} \int_0^{\infty} \epsilon^{1/2} \ln \left(1- z_q e_q^{- \beta \epsilon} \right) d\epsilon \nonumber \\
&=& -\frac{2 \pi V}{h^3} (2m)^{3/2} \Biggl[ \ln \left(1- z_q e_q^{- \beta \epsilon} \right) \frac{\epsilon^{3/2}}{3/2} \bigl\vert_0^{\infty} \nonumber \\ &-& \int_0^{\infty} \frac{z_q \beta \left(e_q^{- \beta \epsilon}\right)^q}{1- z_q e_q^{- \beta \epsilon}} \frac{\epsilon^{3/2}}{3/2} d\epsilon \Biggr]
\eea
For $\epsilon \to 0$, $\epsilon^{3/2}=0$ and for $\epsilon \to \infty$, $\ln \left(1- z_q e_q^{- \beta \epsilon} \right)= \ln (1-0)=0$
\bea
\therefore - \sum_{k=0}^{\infty} \ln \left\lbrace 1- z_q e_q^{- \beta \epsilon_k} \right\rbrace &=& -\frac{2 \pi V}{h^3} (2m)^{3/2} \left[- \int_0^{\infty} \frac{z_q \beta \left(e_q^{- \beta \epsilon}\right)^q}{1- z_q e_q^{- \beta \epsilon}}  \frac{\epsilon^{3/2}}{3/2} d\epsilon \right] \nonumber \\
&=& \frac{2 \pi V}{h^3} (2m)^{3/2} \frac{2 \beta}{3} \int_0^{\infty} \frac{ \epsilon^{3/2}~d\epsilon}{z_q^{-1} \left( e_q^{-x}\right)^{-q}-\left( e_q^{-x}\right)^{1-q}} \nonumber \\  &=& \frac{2 \pi V}{h^3} (2m)^{3/2} \frac{2 \beta}{3} \frac{1}{\beta^{5/2}} \int_0^{\infty} \frac{ x^{3/2}~dx}{z_q^{-1} \left( e_q^{-x}\right)^{-q}-\left( e_q^{-x}\right)^{1-q}} \nonumber \\ &=& \frac{2 \pi V}{h^3} (2m k_B T)^{3/2} \frac{2}{3} \Gamma(5/2) ~ gb_{5/2}(z_q) \nonumber \\ &=& \frac{2 \pi V}{h^3} (2m k_B T)^{3/2} \frac{2}{3} \frac{3}{2} \frac{\sqrt{\pi}}{2} ~ gb_{5/2}(z_q) \nonumber \\
&=& \frac{V}{\lambda^3} ~ gb_{5/2}(z_q)
\eea
where $x=\beta \epsilon$ and $ \lambda=\frac{h}{ \sqrt{2 \pi m k_B T}}$, the de-Broglie wavelength.
\protect\label{sum_to_integral}

\section{Properties of $q$-generalized Bose integrals(in other words $q$-generalized Polylog functions)}
We have introduced the intermediate functions namely as $q$-generalized polylog functions of first kind and second kind for convenience. These are not any new functions but the $q$-generalized version of the known polylog function $g$ or in other words $q$-generalized Bose integrals.

\begin{figure}[htbp]
  \centering
  \includegraphics[width=10cm]{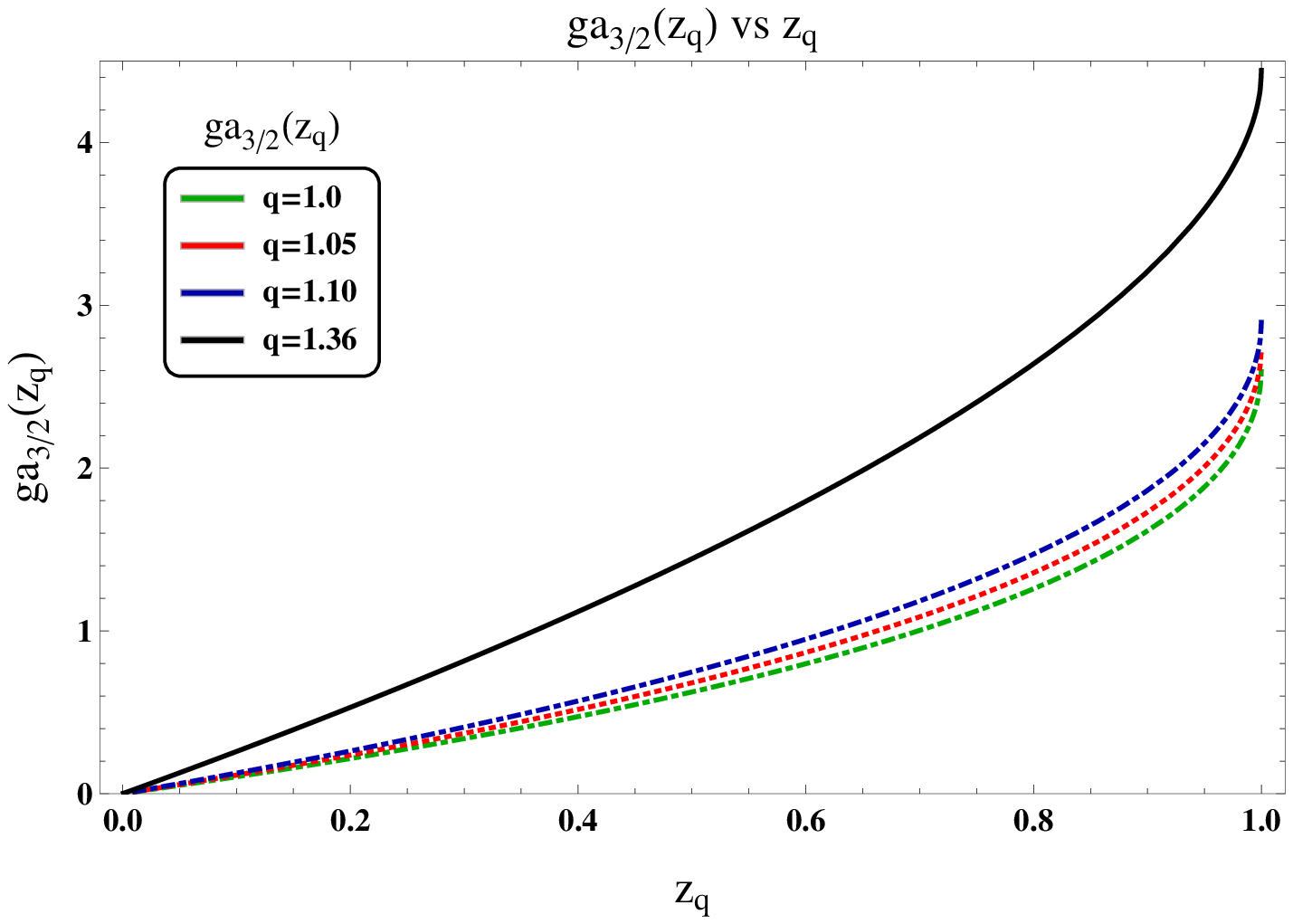}
 \caption{\it Characteristic curve of $ga_{3/2}(z_q)$, $q$-generalized Bose integral of the first kind.} 
\protect\label{q_Bose_integral1}
\end{figure}

 In Fig. \ref{q_Bose_integral1} we have shown the characteristics of $ga_{3/2}(z_q)$ for different $q$. This is the $q$-generalized Bose integral of the first kind, given by
\bea
ga_{3/2}(z_q)=\frac{1}{\Gamma(3/2)} \int_0^{\infty} \frac{dx ~ x^{3/2-1} }{\left( z_q e_q^{-x}\right)^{-q}-1}
\label{q_polylog1}
\eea

\begin{figure}[htbp]
\centering
\includegraphics[width=10cm]{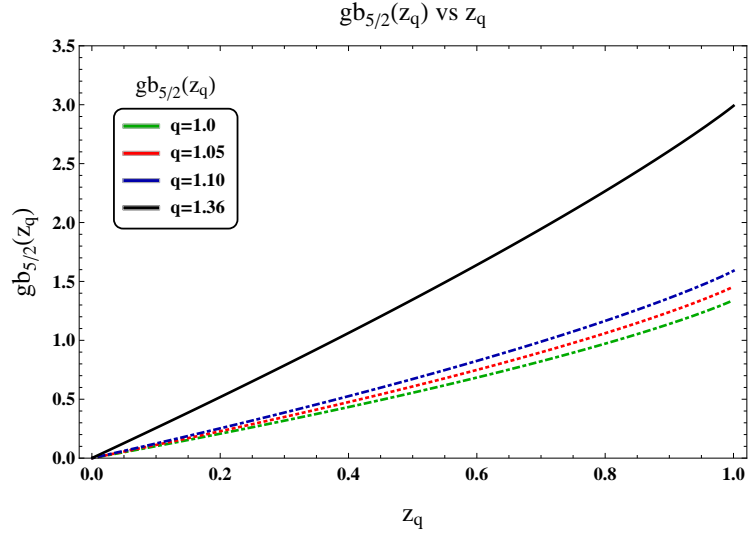}
\caption{\it Characteristic curve of $gb_{5/2}(z_q)$, $q$-generalized Bose integral of the second kind.}
\label{q_Bose_integral2}
\end{figure}

 In Fig. \ref{q_Bose_integral2} we have shown the characteristics of $gb_{3/2}(z_q)$ for different $q$. Here this is the $q$-generalized Bose integral of the second kind, given by
\bea
gb_{5/2}(z_q)=\frac{1}{\Gamma(5/2)} \int_0^{\infty} \frac{dx ~ x^{5/2-1} }{z_q^{-1} \left( e_q^{-x}\right)^{-q}-\left( e_q^{-x}\right)^{1-q}}
\label{q_polylog2}
\eea

 Interestingly both the functions in $q \to 1$ limit i.e., Eqs. (\ref{q_polylog1}) and (\ref{q_polylog2}) become our known Bose integrals which are as follows
\bea
g_{3/2}(z)=\frac{1}{\Gamma(3/2)} \int_0^{\infty} \frac{dx ~ x^{3/2-1} }{\left( z e^{-x}\right)^{-1}-1}
\eea
and
\bea
g_{5/2}(z)=\frac{1}{\Gamma(5/2)} \int_0^{\infty} \frac{dx ~ x^{5/2-1} }{z^{-1}  e^{x}-1}
\eea

 For $z_q=1$ these functions become Riemann-Zeta functions. In general for any $q$, Eq. (\ref{q_polylog1}) is equivalent to the $q$-generalized version of the Bose integrals given in \cite{Ou, Chen}. The plot obtained for the critical temperature $T_c$ as a function of the deformation parameter $q$ using our $q$-modified version of the polylog function of first kind and the same plot using the $q$-generalized version of the Bose integral given by \cite{Ou, Chen}, are overlapping with each other(that is why we could not show the prediction of Ou \etal and Chen \etal in Fig. \ref{tcq} separately \cite{Ou, Chen} with our prediction). The only difference is that, Ou \etal and Chen \etal used the same $q$-generalized version of the Bose integral as well to calculate the specific heat, whereas we used $q$-generalized Bose integral of the second kind to calculate the same. That is why our specific heat characteristics are little bit different from their prediction.
 
 It is very difficult to derive an exact expression for the $q$-generalized mean
occupation numbers from a more fundamental statistical description of a Bose gas which is strictly derived from more basic assumptions. Though we attempted to solve this issue using some algebraic assumptions which are reasonably valid under very restrictive conditions, an alternative strategy has been followed by Ou \etal and Chen \etal \cite{Ou, Chen}. They assumed a reasonably well defined $q$-generalized expression for the occupation number which is compatible with the general structures of the $q$-thermostatistical formalism. They investigated thoroughly and found that it exhibits
physically appealing properties. Most importantly, though that expression cannot be obtained from the first principle at the moment, it agrees with the experimental result. Clearly the sensible choice of such an expression can correctly describe the experimental data.

\end{document}